\documentstyle[preprint,aps,prb]{revtex}

\begin{document}
\draft
\preprint{CIEA-Phys. 02/96}
\title{
Electronic structure of the valence band of II--VI wide band gap 
semiconductor interfaces}

\author{D. Olgu\'{\i}n and R. Baquero}

\address{
Departamento de F\'{\i}sica, Centro de Investigaci\'on y de Estudios 
Avanzados del IPN.,\\ Apartado Postal 14-740, 07000 M\'exico D.F.
}
\maketitle
\begin{abstract}
In previous work we have discussed in detail the electronic band 
structure of a (001) oriented semi--infinite medium formed by some
II--VI zinc blende semiconductor compounds in the valence band 
range of energy. Besides the known bulk bands (hh, lh and spin--orbit
splitting), we found two characteristic surface resonances, one 
corresponding to the anion termination and another to the cation one.
Furthermore, three (001)--surface--induced bulk states with no--dispersion
from $\Gamma-X$ are also characteristic of these systems. 

In this work we present the electronic band structure for (001)--CdTe
interfaces with some other II--VI zinc blende semiconductors. We 
assume ideal interfaces. We use tight binding Hamiltonians with 
an orthogonal basis ($s p^3 s^*$). We make use of the well--known
Surface Green's Function Matching method to calculate the interface 
band structure. In our calculation the dominion of the interface is 
constituted by four atomic layers. We consider here anion--anion
interfaces only. We have included the non common either anion or 
cation (CdTe/ZnSe), common cation (CdTe/CdSe), and common anion 
(CdTe/ZnTe) cases. We have aligned the top of the the valence band 
at the whole interface dominion as the boundary condition.

The overall conclusion is that the interface is a very rich space 
where changes in the band structure with respect to the bulk do 
occur. This is true not only at interfaces with no common atoms 
but also at the ones with either common cation or anion atoms 
irrespective to the fact that the common atomic layers are facing or 
not each other at the interface. 

Finally, we found that the (001)--surface--induced bulks states reappear
at the interface in contrast to the pure (001)--surface resonances 
which disappear. This confirm our previous interpretation of such 
states as {\it bulk} states. Their behaviour is very interesting 
at the interface. We have refine the terminology for these states
to up--date it to the new results and have call them {\it Frontier
induced semi--infinite medium} (FISIM) states. They might well
appear also in quantum wells and superlattices and have influence
in the transport properties of these systems.

\end{abstract}


\narrowtext

\section{Introduction}

In the last ten years, the study of the physics of surfaces, interfaces,
superlattices and quantum wells of semiconductors has been the object of
permanent study.%
\cite{bryant,lowther,kilday,niles,bala,gawlik,duc,arriaga,prb1} At the
origin of the deep understanding of the experimental results on these
systems is an accurate description of its electronic band structures and
its phonon spectra. In previous
papers,\cite{prb1,rafa,noguera,quintanar,rafa-prb,angela,daniel,prb2} in
conjunction with the known surface Green's function matching method
(SGFM)\cite{g-m}, we have used a tight-binding formulation to calculate
the electronic band structure, the surface, the surface induced, and the
interface states for several systems in a consistent way with the known
bulk band structure calculations. The method can also be applied to
overlayers,\cite{martin} superlattices,\cite{arriaga,rafa-moliner}
phonons\cite{brito} and to calculate transport properties \cite{col} in
heterostructures as quantum wells, for example, by making use of the well
known method by Keldysh.\cite{keldysh}

Although there is no actual need to restrict ourselves to
non-reconstructed ideal interfaces, we consider interesting to deal with
this simpler situation. This is not a real limitation. 
 Recent advances in thin film deposition technology\cite{falco} have allowed
the fabrication of overlayers on surfaces, interfaces, and superlattices
under stricter control of the parameters entering in the process of production, 
and samples with high degree of structural coherence are now possible. 
These artificially prepared materials are of great interest since they can 
exhibit properties different from those that occur in nature. Nevertheless, 
a minimum perfection in the growth of the samples is to be achieved 
before the properties manifest themselves. For example, superlattices
will not exhibit their specific electronic properties unless interdiffusion 
between the two media is avoided to a great extent.\cite{falco}

The semiconductor interfaces are less studied than their surfaces.
Although interesting in themselves, the deep understanding of the physics
of the interface is an important starting point for the detailed study of
thermal, optical, photoacoustic and other properties of quantum--wells and
superlattices.  The electronic properties at solid--solid interfaces
depend sometimes even on details of the interaction between the two atomic
layers from the different materials in contact. Our work can be used as a
starting point to analyze those details. These are responsible for the
characteristics of interface reconstruction, thermodynamic properties,
degree of intermixing, stress, compound formation, etc. 

In previous work,\cite{prb1,prb2} we have studied the electronic structure
of the valence band for the (001)--surface of several II--VI wide band gap
semiconductors. We have considered CdTe, ZnTe, CdSe and ZnSe. We have
obtained the bulk bands (infinite medium) from the direct diagonalization
of our tight-binding Hamiltonians and compared our results with the available
data. The same result was obtained using the SGFM method, from the
(001)-bulk-projected Green's function. This is actually a proof of
consistency which gives us confidence in the new results. The main
characteristics of the electronic spectra for these materials appear in
Table I. 

The general characteristic of the zinc blende II--VI wide band gap
semiconductor valence band as we have obtained from our calculations is
sketched in Fig. 1. The hh and lh bands follow each other closely in
energy. The hh band disperses from $\Gamma$ to $X$ about 2.0 eV and the lh
one about 2.4 eV. The spin-orbit splitting is about 1.0 eV in the
Te--compounds and about 0.5 eV in the Se--ones. The spin--orbit band (SO)
reaches $X$ at about 5.0 eV. This is the band with the most dispersion.
The LCAO composition is mainly ($p_x,p_y$) for the first two bands and
$(p_z)$ for the last one. 

From the (001)-bulk-projected Green's function we get the energy of the
(001)-surface-induced bulk states. Three such states appear, $B_h$, $B_l$
and $B_s$, in this range of energy. These surface-induced bulk bands show
no dispersion\cite{prb1,daniel,prb2} and were first found experimentally
by Niles and H\"ochst\cite{niles} and confirmed later by Gawlik {\it et
al.}\cite{gawlik} for CdTe(001). At $X$ $B_h$ mixes with the hh bulk band
and $B_l$ with the lh one. Both are of ($p_x,p_y$)--character. $B_s$ mixes
with the spin-orbit band at $X$ and is mainly ($s,p_z$). The three states
appear at the same position in energy irrespective of the cation or anion
termination of the surface as one expects for surface-induced {\it bulk}
states which only depend on the surface through the boundary condition
(the wave function has to be zero at the surface). 

The (001)-surface valence band is very rich\cite{prb1,prb2} in several
other features. In particular, two characteristic surface states do exist
in this range of energy. One corresponds to the anion ($S_a$) and the
other to the cation ($S_c$) termination of the (001)-surface. In all the
systems considered, the anion terminated surface band follows roughly the
dispersion of the heavy hole bulk band but is at a slightly higher energy.
The cation terminated surface band starts roughly around 2--3 eV from the
top of the valence band in $\Gamma$ and has a varying amount of
dispersion. The two states appear at very different energy values and are
distinctive of the termination of the surface for the four systems under
consideration. 

In this work, we want to extend these findings to the case of interfaces.
We will study the valence band of several (001)-CdTe interfaces with other
II--VI wide band gap zinc blende semiconductors, namely, (001)--CdSe,
(001)--ZnTe and (001)--ZnSe. Our work can be straightforwardly extended to
other interfaces using the data that we give in the appendix. Our paper is
organized as follows. In section II, we summarize the method that we have
used, section III is devoted to discuss our results and constitutes the
main part of the paper. We describe in detail the valence band of the
interfaces studied in this section. 
Section IV is devoted to the discussion of the frontier induced semi--infinite
medium (FISIM) states which are reminiscent of the surface--induced 
bulk states.
We then summarize our conclusions
in a final section V.

\section{Method}

To describe the interface between two semiconductor compounds, we make use
of tight--binding Hamiltonians. The Green's function matching method takes
into account the perturbation caused by the surface or interface exactly,
at least in principle, and we can use the bulk tight--binding parameters
(TBP).\cite{rafa,noguera,quintanar} This does not mean that we are using
the same TBP for the surface, or for the interface and the bulk.  Their
difference is taken into account through the matching of the Green's
functions. We use the method in the form cast by Garc\'\i a--Moliner and
Velasco.\cite{g-m} They make use of the transfer matrix approach first
introduced by Falicov and Yndurain.\cite{falicov} This approach became
very useful due to the quickly converging algorithms of L\'opez--Sancho
{\it et al.}\cite{sancho} Following the suggestions of these authors, the
algorithms for all transfer matrices needed to deal with these systems can
be found in a straightforward way.\cite{trieste} This method has been
employed successfully for the description of
surfaces,\cite{prb1,rafa,noguera} interfaces,\cite{quintanar,rafa-prb} and
superlattices.\cite{arriaga,rafa-moliner}

\subsection{The Formalism}

We have calculated\cite{prb2} the bulk (infinite medium) band structure of
the compounds by the tight--binding method (TB) in the Slater--Koster
language\cite{slater} using an orthogonal basis of five orbitals,
$sp^3s^*$. The $s^*$ state is introduced to properly locate in energy the
conduction band usually formed by $d$ states in the II--VI zinc--blende
(ZB) semiconductor compounds.\cite{austria,harrison} We have included the
effect of the spin--orbit (SO) interaction.\cite{chadi} The TBP that we
have used are listed in the Appendix. They reproduce the known bulk band
structure calculations for all the semiconductors studied
here.\cite{prb1,bertho,ang-cdse}

To calculate the Green's function for the interface, we first get the one
for the free surface. We assume ideal truncation.  The general equation
for the Green's function can be written as: 
\begin{equation}
\label{uno}
(\omega I-H)G=I, 
\end{equation}
where $\omega$ is the energy eigenvalue and $I$ is the unit matrix. We
adopt the customary description in terms of principal layers.  Let $\mid
n\rangle$ be the principal wave function describing the $n^{\rm th}$
principal layer. It is a LCAO wave function formed by one $s$--like, three
$p-$like, and one $s^*$--like atomic wave function per spin per atom.
Since there are two atomic layers in a principal layer, $\mid n\rangle$ is
a 20--dimensional vector (2 spin states). If we take matrix elements of
eq. (\ref{uno}) in the Hilbert space generated by the complete set of wave
functions $\mid n\rangle$, we get 
\begin{equation} \label{dos} \langle n
\mid(\omega I-H)G\mid m \rangle = \delta_{mn}. 
\end{equation}

Since, by definition, only nearest--neighbor interactions take place 
between principal layers, the identity operator 
to be introduced between $(\omega I-H)$ and $G$ can be 
expressed as:
\begin{equation}
\label{tres}
I=\mid n-1\rangle\langle n-1\mid + \mid n\rangle\langle n \mid +
\mid n+1n\rangle\langle n+1\mid .
\end{equation}
By inserting            
(\ref{tres}) 
into (\ref{dos}) we get 
\begin{equation}
\label{cuatro}
(\omega -H_{nn})G_{nm}-H_{nn-1}G_{n-1m}-H_{nn+1}G_{n+1m}=
\delta_{mn}.
\end{equation}
This is because  $H_{m,m+i}=0$ for $i\geq2$. 
The matrix elements of the Hamiltonian, $H_{nm}$, that appear in this 
formula are 2$\times$2 supermatrices. 
For example, in the case of a surface
\begin{mathletters}
\begin{equation}
H_{00}=\left(
\begin{array}{lr}
	h_{00} & h_{0-1} \\
	h_{-10}& h_{-1-1} \\
\end{array}\right),
\end{equation}
\begin{equation}
H_{01}=\left(
\begin{array}{lr}
	h_{0-2} & h_{0-3} \\
	h_{-1-2}& h_{-1-3} \\
\end{array}\right).
\end{equation}
\end{mathletters}
Notice that rows are labeled with the index of the surface principal
layer zero (containing atomic layers 0 and --1, for both $H_{00}$ and 
$H_{01}$) while the columns are 
indexed with the zero and first principal layer (atomic layers 
0 and --1, and --2 and --3, for $H_{00}$ and $H_{01}$, respectively). We 
label principal layers with positive numbers and atomic layers with 
negative numbers. The surface is labeled with zero in both cases. We 
shall adopt the hypothesis of an ideal, non reconstructed surface. 
For the (001)--surface of a II--VI compound we have one atomic layer
of anions and one of cations per principal layer. 
In this case $h_{00}\neq h_{-1-1}$ but $h_{0-1}=h_{-10}^\dagger$. 
To calculate $H_{00}$ and $H_{01}$ we only need to know $h_{00},\ h_{-1-1},\ 
h_{0-1},$ and $h_{-1-2}$, since $h_{0-2}=h_{0-3}=0$ in the first nearest 
neighbors approximation.
These matrices are readily written in the 
tight--binding language and can be calculated with {\it the bulk} TBP 
as mentioned above. All the $h-$matrices are functions of the wave 
vector {\bf k}.

Using (\ref{cuatro}) for $n=m$, and $m=0$ for the surface, it is 
straightforward to get the surface Green's function\cite{g-m}
\begin{equation}
\label{seis}
G_s^{-1}=(\omega I -H_{00})-H_{01}T
\end{equation}
and the principal layer projected bulk Green's function\cite{g-m}
\begin{equation}
\label{siete}
G_b^{-1}=G_s^{-1}-H_{01}^\dagger \widetilde T.
\end{equation}

It is customary to define the transfer matrices as
\begin{mathletters}
\begin{equation}
G_{k+1p}=TG_{kp},  \qquad k\geq p\geq 0
\end{equation}
\begin{equation}
G_{ij+1}=\widetilde TG_{ij},  \qquad j\geq i\geq 0.
\end{equation}
\end{mathletters}
These matrices can be calculated by the quick algorithm of L\'opez--Sancho 
{\it et al.}\cite{sancho}
(see also Refs. [11, 12, 26] for a compilation of
the formulae and details of the algorithms used).

In the case of the interfaces the matrices double in size. The algebra is 
the same. One gets
\begin{equation}
\label{infcs}
G_I^{-1}=G_{s(A)}^{-1} + G_{s(B)}^{-1} - I_BH^iI_A - I_AH^iI_B,
\end{equation}
which is the analogous formula to (\ref{seis}) above. $G_I^{-1}$ is the 
interface Green's function, $G_{s(A)}^{-1}$ and $G_{s(B)}^{-1}$ are 
the surface Green's function of medium A and B, respectively, in the 
doubled space of the interface,
\begin{equation}
\label{diez}
G_{s(A)}^{-1}=
\left(
\begin{array}{lr}
		G_A^{-1} & 0\\
		0        & 0
\end{array}\right)
\end{equation}
where $G_{A}^{-1}$ is the surface Green's function for the medium $A$
calculated from (\ref{seis}). It is a $20\times20$ matrix while 
$G_I^{-1}$ 
is a $40\times40$ matrix. $I_BH^iI_A$ and $I_BH^iI_B$ are $40\times40$
matrices of the form
\begin{equation}
-I_BH^iI_A - I_AH^iI_B =
\left(
\begin{array}{lr}
		0  &  -{\cal I}_AH^i{\cal I}_B \\
	  -{\cal I}_BH^i{\cal I}_A &   0
\end{array} \right)
\end{equation}
they describe the interaction between the two media. 
$-{\cal I}_AH^i{\cal I}_B$
and $-{\cal I}_BH^i{\cal I}_A$ are $20\times20$ matrices. In our
model they take the form of the surface Hamiltonian $H_{01}$ and $H_{01}^
\dagger$, respectively, with TBP taken as the average of those for the two 
media. This is a reasonable approximation when both sides of the interface
have the same crystallographic structure and we take the same basis of 
wave functions. 

From the knowledge of the Green's function, the local density of states
can be calculated from its imaginary part integrating over the
two--dimensional first Brillouin zone. We have applied previously this
formalism to surfaces\cite{rafa,noguera}, interfaces%
\cite{quintanar,rafa-prb} and superlattices\cite{rafa-moliner}. Now we
present our results. 

\section{Results and discussion}

This section is devoted to the discussion of the interface valence band of
the (001)--projected electronic band structure of II--VI zinc blende wide
band gap semiconductors. We have calculated all the interfaces formed by
CdTe, ZnTe, CdSe and ZnSe. In Table I we present the main characteristics
of the band structure for these materials, as we already mentioned. We
will present in this paper the CdTe--(001) interfaces in detail (see Figs.
2--4 and Table II) and we give all the necessary data so that the full
electronic band structure for the rest of the interfaces can be
reproduced, as experiments become available. 

The domain of the interface is constituted by several atomic layers. Since
we are considering first nearest neighbors interactions in our bulk
Hamiltonians, we will consider four atomic layers as the interface domain,
two belonging to medium A and two to medium B. To distinguish between the
different atomic layers we will call each atomic layer by the medium its
neighbors belong to. The atomic layer AA will be the second from the
interface into medium A.  AB will be the last atomic layer belonging to
medium A and facing the first atomic layer of medium B and so on. 
So the four atomic
layers that constitute the interface domain will be labeled AA, AB, BA,
and BB.  For the interfaces aligned along the (001) direction the two
media are facing each other either through its anion or cation atomic
layer.  We will consider
here the case of an anion-anion interface only. The notation
(001)-CdTe/ZnSe means an interface constituted by two zinc blende crystals,
CdTe and ZnSe, both aligned along the (001)-direction and where the atomic
planes are identified as follows AA=Cd, AB=Te, BA=Se and BB=Zn. We will
project the interface electronic band structure on each atomic layer and
see how the different states that we found for the free surface case
change or disappear at the interface.  New states, in principle, can
appear at the interface. These interface states are important to known
since they may play a role in the transport properties of
heterostructures. 

One important point is the band offset used. Experimental results for some
II--VI semiconductors are available.\cite{duc,pelhos} In general, for
II--VI semiconductors, the anion-anion interfaces have small valence
band--offsets and cation--cation ones have small conduction
band--offsets (both of the order of some meV). The rest of the bands and
the surface--induced bulk states will show discontinuities accordingly. We
will use the boundary condition that the top of the valence bands at the
interface are aligned and choose this energy as our zero. As a consequence
the conduction band offset will be equal to the difference in the band
gaps. The surface-induced bulk states (see Table III) do survive at the
interface; their energy position changes slightly and show dispersion in
some cases contrary to the semi--infinite case (see Figs. 2--4 and Table
IV). The interface builts up discontinuities characteristic of each one of
them (see text below).  The surface states do not survive and are not
present at the interface domain, as expected. This gives further support
to our interpretation of the --4.4 eV non--dispersive state\cite{prb1}
found experimentally in CdTe--(001) by Niles and H\"oscht\cite{niles} and
by Gawlik {\it et al.}\cite{gawlik} as a {\it bulk} state. We do not find
interface resonances in this interval of energy. 

In Figs. 2--4 and Tables II--IV, we show the electronic band structure of
the valence band for the interfaces studied here, (001)--CdTe/ZnSe,
(001)--CdTe/CdSe, and (001)--CdTe/ZnTe. The dispersion relations are found
from the poles (symbols in the figures) of the real part of the interface
Green's function. The solid--lines are a guide to the eye. These are to be
compared to the dispersion curves found for the bulk (infinite medium)
case. As we mentioned above, the (001) surface-induced {\bf bulk} states
do survive as expected (stars, crosses and dots in Figs. 2--4). We should
refine the convention for {\it surface-induced} bulk states as a result of
these new findings and call them rather {\it frontier--induced
semi--infinite medium} (FISIM) states to incorporate the interface case
and probably other type of heterojunctions as well, and to introduce
clearly the idea that they are semi--infinite medium states as opposed to
bulk (infinite medium) or surface states. In the semi--infinite
medium\cite{prb1,prb2} case we have called these states $B_h$, $B_l$, and
$B_s$ according to the bulk band they mixed with at $X$ in the First
Brillouin zone. We now proceed to give a more detailed account for each
particular interface studied. 

\subsection{The (001)-CdTe/ZnSe interface}

Fig. 2 shows the electronic structure of the valence band (VB) for this
interface. We consider an anion--anion (Te--Se) interface. As boundary
condition we have aligned the top of the VB projected onto the atomic
layer Te (AB) with the one projected onto the atomic layer Se (BA). We
take this common energy value as our zero of energy. All the energies are
in electron--volts (eV). The value of the top of the valence band
projected onto the AA(Cd)-- and BB(Zn)-- atomic layer is for both also
0.0. There is a conduction band (CB) discontinuity at the interface of
about 1.2 at the anion--anion interface (see Table I). 

In Fig. 2a, we show the electronic band structure projected onto the Cd
(AA) atomic layer. The full lines are the heavy hole (hh), light hole (lh)
and the spin--orbit (SO) projected dispersion curves. They are actually a
fit to the computed points which appear as empty triangles. In Table II,
we summarize the main characteristics of the electronic dispersion curves
for the interfaces under consideration. When compared to the bulk bands,
the first difference with respect to the interface bands, is the larger
width at the interface for each individual band although the total width
for the three bands shown in the figure remains unchanged. The hh and lh
band width in the bulk are 1.7 and 2.2 (see Table I), respectively,
while at the AA(Cd) atomic layer at the interface are 2.1 and 2.3,
respectively (see Table II). The degeneracy at this point is not broken by
the interface.  For the SO--shift in $\Gamma$ the bulk and the surface
band structure give both 0.95 (see Table I) in agreement with experiment.
At the AA(Cd)--atomic layer of the interface, we find a smaller value,
0.6, as it is shown in Table II. This difference should be accessible to
experimental verification. The SO--curve band width is again higher at the
interface (3.8) than at the surface or in the bulk (3.5) as it is
seen from a comparison of the corresponding values in Tables I and II.
Nevertheless the SO--curve projected onto the AA(Cd)--atomic plane
reaches the $X$ high--symmetry point of the Brillouin zone at --4.4, and
the total width of the three bands is the same in both cases (see Fig. 2a). 

Another interesting aspect of the electronic band structure at the
interface is the behaviour of the FISIM--states. We recall that the
(001)--surface induces non-dispersion states\cite{niles,gawlik,prb1,prb2}
in zinc blende II--VI semiconductors. In Figs. 2--4, the straight dotted
lines are the dispersion curves found for the (001) surface case for the
FISIM--states. These states do have dispersion at the interface domain as
a consequence of the influence of the other crystal atoms instead of the
vacuum existing at the surface frontier. 

In Figs. 2, these states are labeled $B_{Ih},\ B_{Il}$ and $B_{Is}$,
reminiscent of the corresponding notation used for the surface--induced
ones (see also Table III). The dotted lines show the FISIM--states
dispersion curves for the (001)--surface--induced states in the binary
crystal case (CdTe or ZnSe). Fig. 2a (AA = Cd) shows that the $B_{Ih}$
(stars) begins at --2.1 in $\Gamma$ and ends in $X$ at about --2.2. As in
the free (001)--surface case, this state is degenerate in $\Gamma$ and
bifurcates in two branches with an overall dispersion of about 0.5. 
$B_{Il}$ (crosses) shows less dispersion (--3.0 in $\Gamma$ 
and --3.3 in $X$). The
energy difference is 0.3. In contrast $B_{Is}$ (dots, at --4.7 and non
degenerate as well) shows almost no dispersion. See Fig. 2a and Table IV.
The energies at $\Gamma$ and at $X$ are, in general, slightly different
from the corresponding ones in the (001)--surface case, see Table III. 
Fig. 2b shows the
projection of the electronic band structure onto the (AB=) Te--atomic
layer.  A very similar analysis with analogous conclusions is made in this
case. It is to be noticed that the FISIM--states show much more dispersion
for this atomic layer projection than before. Also the overall width is
larger than in the bulk case (see Table IV). 

The other side of the interface is presented in Figs. 2c and 2d where the
projection onto the Se-- and Zn--atomic layer is shown, respectively. The
top of the VB is fixed at 0.0 in the Se--projected dispersion relation.
This constitutes the boundary condition. From Table I, we see that for
ZnSe the bulk and the free (001) surface hh band takes the value of --1.95
at the $X$ high-symmetry point. This value should be compared with --2.0
in the interface region according to Figs. 2 and Table II. The lh state in
the free (001) surface takes at $X$ almost the same value for CdTe and
ZnSe, namely --2.2. The same value is found at the ZnSe side of the
interface; see Tables I and II. A notable difference can be observed in
the SO--band projected on the Zn--atomic layer. For this band we obtain
the value of --4.8 at $X$ at the interface domain which is quite less than
the free (001)--surface value of --5.3 (see Table I). 

From Table III we see that for the free ZnSe (001)--surface $B_h$ is at
--2.0.  The corresponding FISIM--state (see Fig. 2d) at the interface,
$B_{Ih}$, shows a slight dispersion of about 0.5 in the $\Gamma-X$
region. $B_l$ in the free (001)--surface is located roughly at --2.2 for
both materials and at the interface ($B_{Il}$) it is at --2.4. Finally, the
$B_{Is}$ state at the BA(Se) atomic layer projection of the interface
follows roughly the non dispersive property of the corresponding
(001)--surface FISIM state. The BB(Zn) projection shows a small
dispersion. 

In general, these FISIM--states are located at different energies in
different materials and are either follow the non dispersive property of
the (001)--surface ones or move to slightly deeper values in energy at
the $X$ high--symmetry point of the interface, as can be checked from
Tables III and IV and from Figs. 2. 
 
In conclusion, the VB of (001)--CdTe/ZnSe interface does present the
overall features of the free (001)--surface ones but at energies that are
slightly different. As it is expected the surface resonances do not appear
at the interface. The FISIM--states do appear, a fact that is also
expected since they are semi--infinite medium states. These states merge
to deeper energies at this interface. They are mainly of $s-p$ anion
states character. In Table V, we give the LCAO--composition for all the
states that appear at the interfaces studied. The valence band offset used
is zero.

\subsection{The (001)-CdTe/CdSe interface}

Here again, as we explained before, our notation means two zinc blende
crystals aligned in the (001) direction forming an interface. In our
previous notation AA=Cd, AB=Te, BA=Se, and BB=Cd. The projected band
structure is presented in Fig. 3. The symbols in the figure are obtained
from the poles of the real part of the interface Green's function obtained
within the SGFM formalism as before. The lines are interpolations meant as
a guide to the eye. These should be compare to the bulk bands. The bulk
band parameters of interest here for the two crystals forming the
interface are quoted in Table I. 

We consider an anion--anion (Te--Se) interface as before and the boundary
condition is to align the top of the valence band in both sides of the
interface. We take this common value as our zero of energy which we
express again in eV. As it can be seen from Table I, the two crystals
forming this interface have the smallest difference in their bulk gap
value (0.18) in contrast with the previously studied (001)-CdTe/ZnSe where
it is the highest (1.2). As a consequence of our used boundary condition
this is the CB offset. This is therefore the interface with the smallest
CB offset studied here. 

The first thing to compare is the behaviour of the bands in the common
medium A(CdTe) for the previous and this interface. In the AA(Cd)
projection the hh band behaves very similar. It is only the value at the
$X$--high symmetry point that differs by 0.3 (Table II and Figs. 2a and
3a). In contrast the lh bands develop differently in the sense that some
points are at a higher distance in energy from the respective hh band
ones in both graphs. In the
AA(Cd) projection the lh band appears to be the more affected by the
presence of a different partner at the interface. 
The SO-curve is again similar in both cases. The {\it
Frontier-induced semi-infinite medium} (FISIM)--states associated with the
lh and SO bands behave very similarly but, in contrast, the one associated
with the hh bulk band differs for the AA(Cd) projection at the two
interfaces. 

The anion AB(Te) atomic layer projected bands do differ but slightly for
the two interfaces. 

Next, it is interesting to compare the behaviour of the AA(Cd) projected
bands with the BB(Cd) ones for this (001)--CdTe(A)/CdSe(B) anion--anion
interface. The hh and lh curves for the BB(Cd) projected bands at this
interface (Fig. 3d) behave very similar to the AA(Cd) (Fig. 2a) ones for
the (001)--CdTe/ZnSe interface but the SO curves do not. The FISIM--states
do differ in the way they disperse and in the energy position value. 

This common--cation anion--anion interface presents also the interesting
feature that the spin-orbit shift is different at the AB(Te) and BA(Se)
atomic layer projections (0.8 and 0.9, see Table II and Figs. 3b, and 3c) 
than at
the AA(Cd) and BB(Cd) projections (0.5, Table II and Figs. 3a and 3d). 

In conclusion, the band structure for this interface shows that the very
same atomic layer not facing directly the other medium at the interface
can present noticeably differences in the projected band structure in
different interfaces and in different sides of the same interface. This
introduces more subtleties into the complicated problem of the
calculation of the band offsets. 

\subsection{The (001)-CdTe/ZnTe interface}

This is a different case, the common--anion anion--anion interface. We use
again the same boundary condition for the VBs. The CB offset is then equal
to the difference between the bulk band gaps. One interesting point in
this case is to compare the AB(Te) and BA(Te) projected band structures
(See Figs. 4b and 4c and Tables II and IV). Another point of interest is
to compare the behaviour of the band structure at the BB(Zn) atomic layer
for this interface with the BB(Zn) for the (001)--CdTe/ZnSe interface
studied previously (See Figs. 2d and 4d and Tables II and IV). The
analysis is very similar to the one made above and leads to the new
conclusion that the same atomic layer (Te) facing each other at an
interface can show characteristics of the band structure that are
different. 

\section{The FISIM--states}

The {\it Frontier Induced Semi-Infinite Medium} (FISIM)--states were
reported experimentally by Niles and H\"ochst \cite{niles} and confirmed
later on by Gawlik {\it et al.} \cite{gawlik} in the (001)-CdTe oriented
semi-infinite crystals. We succeeded in showing where these states come
from. \cite{prb1,prb2} We have reproduce their energy position and
non-dispersive character from the poles of the real part of the
semi-infinite medium Green's function. We showed that these states are of
{\it semi-infinite medium (bulk)} character as opposed to {\it surface}
character. The fact that no surface resonances survive at the interface
but the FISIM--states do, confirms our interpretation. The FISIM--states
have some dispersion at the interface domain and it is to be assumed that
the second medium is responsible for the changes in the dispersion
properties of these states. It is to be expected from quantum mechanics
that given a Hamiltonian the changes in the boundary conditions from
infinite--medium to semi--infinite do manifest themselves in changes of
some of the properties of the eigenfunctions in both cases. The
semi--infinite medium wave function losses in a certain direction the
periodic condition. This is essentially the origin of the FISIM states. 

\section{Conclusions}

We have presented the results for three of a series of band structure
calculations for interfaces formed by II-VI wide band gap zinc blende
semiconductors. We have included the non common either anion or
cation, common cation and common anion cases. All our interfaces are
anion--anion ones. The boundary condition used was to align the top of the
valence bands at the interface domain. In our approximation it amounts to
four atomic layers altogether. We have used the known SGFM method to
calculate the interface Green's function and have produced the projected
band structure at each atomic layer of the interface. We have used
tight--binding Hamiltonians to describe the two media. Our parameters give
very reasonable bulk band structures that agree with the known
experimental data for the II-VI wide band gap semiconductors that we have
been concerned with in this work. The parameters are given in the Appendix
so that these and other results can be reproduced by the interested reader.
We recall that the input to SGFM method are the bulk parameters. This does
not mean that the same parameters are used for the surface or the
interface and the bulk. The difference is taken into account by the method
itself. To describe the interaction at the interface we have used linear
combination of the tight-binding parameters for the two media. 

The overall conclusion is that the interface is a very rich space where
changes in the band structure with respect to the bulk do occur. This is
true not only at interfaces with no common atoms but also at the ones with
either common cation or anion atoms irrespective to the fact that the
common atomic layer are facing or not each other at the interface. This
introduces further subtleties to the complicated problem of the
calculation of the band offset. 

A very interesting point is the behaviour of the {\it Frontier Induced
Semi Infinite Medium} (FISIM)--states.  In this work we showed that no
surface resonances survive at an interface but, in contrast, the
FISIM--states do. This confirms the interpretation that we gave before
about these states in the sense that these are not surface states. The
FISIM--states have no dispersion at a semi-infinite medium with a free
surface but show some at an interface (see Figs 2--4 and Table IV). Their
LCAO--composition is shown in Table V for completeness. 

All these details of the interface band structure should manifest in the
transport properties of heterojunctions and should influence the behaviour
of a device. In this concern it will be interesting to study the the band
structure of ternary and quaternary compounds interfaces, quantum wells
and superlattices.

\begin{figure}
\caption{The general characteristic of the zinc blende wide band gap
semiconductor electronic valence band structure. We show the heavy hole
(hh), light hole (lh), spin--orbit (SO), and the lower $s$--cation
character ($b_{10}$) bulk bands. Furthermore, we have the bulk  
surface--induced states $B_h,\ B_l$, and $B_s$ 
(the FISIM states, see text) and the surface resonances $S_{a1},\
S_{a2}$, and $S_c$. For details see Ref. [16].}
\end{figure}

\begin{figure}
\caption{The full electronic valence band structure for (001)--CdTe/ZnSe
interface in the $\Gamma-X$ direction. The dispersion relations
(triangles) are found from the poles of the real part of the interface
Green's function. The solid lines are a guide to the eye. The
FISIM--states are shown as stars, crosses and points. The dotted lines
show the FISIM--states (see text) for the (001)--surfaces in the binary
crystal (from Ref. [16]).}
\end{figure}

\begin{figure}
\caption{Electronic valence band structure of the (001)--CdTe/CdSe
interface. The conventions are the same as in Fig. 2 (see figure caption).}
\end{figure}

\begin{figure}
\caption{Electronic valence band structure of the (001)--CdTe/ZnTe
interface. See figure caption 2 for details.}
\end{figure}

\narrowtext
\begin{table}
\caption{
Characteristic values taken from the bulk electronic band structure for
the four binary II--VI compounds of interest in this work.  The origin is
at the top of the valence band and the energies are given in eV. The
columns give the gap $(\Gamma_6^c)$, the spin--orbit splitting
$(\Gamma_7^v)$, the value of the heavy hole dispersion curve at the $X$
point $(X_7^v)$, the light--hole one $(X_6^v)$ and the spin--orbit one
$(X_6^v)$.}

\begin{tabular}{||c|cc|ccc||}
  & $\Gamma_6^c$ & $\Gamma_7^v$ & $X_7^v$ & $X_6^v$ & $X_6^v$ \\
\tableline
CdTe& 1.602 & --0.9 & --1.7 & --2.2 & --4.4 \\
CdSe& 1.78 &  --0.4  & --2.18& --2.36&--4.89 \\
ZnTe& 2.39 & --0.91 & --1.93 & --2.40 & --5.50 \\
ZnSe& 2.82 &  --0.45 & --1.95 & --2.19 & --5.30 \\
\end{tabular}
\label{tabla1}
\end{table}

\begin{table}
\caption{
Energies for the heavy hole (hh), light hole (lh), and spin--orbit (SO)
bands at $\Gamma$ and $X$ high--symmetry points as obtained for the
interface domain for (001)--CdTe/ZnSe, (001)--CdTe/CdSe, and
(001)--CdTe/ZnTe. The energies are in eV.}

\begin{center}
\begin{tabular}{||c|c|c|ccc||}
 &  & $\Gamma$--point &  & $X$--point &  \\
\hline
\hline
System & Atomic layer & $E_{so}$ & $E_{hh}$ & $E_{lh}$ & $E_{so}$ \\
\hline
            & Cd & --0.6 & --2.1 & --2.3 & --4.4  \\
            & Te & --0.6 & --2.0 & --2.4 & --4.8  \\
CdTe / ZnSe & Se & --0.6 & --2.0 & --2.2 & --5.2 \\
            & Zn & --0.6 & --2.0 & --2.2 & --4.8  \\
\hline
            & Cd & --0.5 & --2.4 & --2.5 & --4.3  \\ 
            & Te & --0.8 & --2.2 & --2.3 & --4.7  \\
CdTe / CdSe & Se & --0.9 & --2.3 & --2.6 & --4.9  \\
	    & Cd & --0.5 & --2.2 & --2.4 & --4.9  \\
\hline
	    & Cd & --1.0 & --2.0 & --2.3 & --4.3  \\
            & Te & --0.8 & --1.7 & --1.9 & --4.5  \\
CdTe / ZnTe & Te & --1.0 & --1.9 & --2.1 & --5.6  \\
            & Zn & --1.0 & --1.9 & --2.1 & --4.8  \\
\end{tabular}
\end{center}
\label{tabla2}
\end{table}

\begin{table}
\caption{
Energies for the FISIM states (see  text) for the free (001)--surface.
Values taken from Ref. [16].}
\begin{tabular}{||c|cccc||}
State  & CdTe & CdSe & ZnTe & ZnSe \\
\hline
$B_h$   & --1.8  & --2.2  & --2.0  & --2.0 \\
$B_l$   & --2.2  & --2.45 & --2.5  & --2.3 \\
$B_s$   & --4.4  & --5.0  & --5.5  & --5.3 \\
\end{tabular}
\label{tabla3}
\end{table}

\begin{table}
\caption{
Energies for the FISIM states (see text), $B_{Ih},\ B_{Il}$, and $B_{Is}$,
at the interface dominion of
(001)--CdTe/ZnSe, (001)--CdTe/CdSe, and (001)--CdTe/ZnTe.}
\begin{center}
\begin{tabular}{||c|c|ccc|ccc||}
    &	&   &  $\Gamma$--point  &    &   &  $X$--point  &     \\
\hline
\hline
System & Atomic layer & $B_{Ih}$  & $B_{Il}$ & $B_{Is}$ & $B_{Ih}$ & $B_{Il}$
							    & $B_{Is}$ \\
\hline
	  & Cd      & --2.1 & --3.0  & --4.7   & --2.2 & --3.3 & --4.8  \\
	  & Te      & --1.5 & --2.5  & --4.8   & --2.4 & --2.6 & --4.8  \\
CdTe/ZnSe & Se      & --1.5 & --2.4  & --5.2   & --2.2 & --2.5 & --5.4  \\
	  & Zn      & --1.9 & --2.4  & --5.2   & --2.2 & --2.5 & --5.4  \\
\hline
	  & Cd      & --2.5  & --2.9 & --4.6   & --2.5 &  --3.2 & --4.8 \\ 
	  & Te      & --2.4  & --2.8 & --4.8   & --2.4 &  --3.0 & --4.7 \\
CdTe/CdSe & Se      & --1.9  & --2.5 & --5.0   & --2.3 &  --2.6 & --5.1 \\
	  & Cd      & --2.0  & --2.5 & --4.9   & --2.4 &  --2.6 & --5.1 \\
\hline
	  & Cd      & --2.0  & --2.9 & --4.5 & --2.3 & --3.2 & --4.8 \\
	  & Te      & --1.8  & --2.7 & --4.7 & --2.3 & --2.7 & --4.5 \\
CdTe/ZnTe & Te      & --2.0  & --2.5 & --5.8 & --2.4 & --2.6 & --5.9 \\
	  & Zn      & --1.9  & --2.4 & --5.8 & --2.1 & --2.8 & --5.8 \\
\end{tabular}
\end{center}
\label{tabla4}
\end{table}

\begin{table}
\caption{ LCAO composition of the wave function for the states existing at
the interface dominion in all the examples considered in this work.}

\begin{center}
\begin{tabular}{||cc||}
State   & Composition \\
\hline
hh      & ($p_x,p_y$) \\
lh      & ($p_x,p_y$) \\
so      & ($p_z$) \\
\hline
$B_{Ih}$   & ($p_x,p_y$) \\
$B_{Il}$   & ($p_x,p_y$)   \\
$B_{Is}$   & ($s,p_z$)  \\
\end{tabular}
\label{tabla5}
\end{center}
\end{table}

\newpage
\appendix
\section*{}

\begin{table}
\caption{ Tight--binding parameters used in our calculation. We used the
notation of Bertho {\it et al.} Ref. [31]}

\begin{tabular}{||c|cccc||}
 Parameter&CdTe$^{\rm a}$&ZnTe$^{\rm b}$& ZnSe$^{\rm b}$& CdSe$^{\rm c}$\\
\tableline
\tableline
	$E_s^a$&       -8.1921 &-9.19000 & -12.42728 & -10.16740\\
	$E_p^a$&        0.3279 & 0.62682 & 1.78236 & 1.03400 \\
	$E_s^c$&       -0.95   & -1.42 & 0.04728 & 1.07977   \\ 
	$E_p^c$&       6.9379  & 3.77952 & 5.52031 & 7.64650  \\
	$V_{ss}$&      -5.0000 & -6.64227 & -6.50203 & -2.89240 \\
	$V_{xx}$&       2.13600& 1.94039 & 3.30861 & 3.01320  \\
	$V_{xy}$&       5.28170& 4.07748 & 5.41204 & 5.73040    \\
	$V_{sp}^{ac}$& 3.31200 & 5.92472 & 1.13681 & 2.16040 \\
	$V_{sp}^{ca}$& 3.63824 & 4.67265 & 5.80232 & 5.65560 \\
	$E_{s^*}^a$  &10.44540& 6.22682 & 7.84986 & 6.02650  \\
	$E_{s^*}^c$  & 6.62960& 6.77952 & 8.52031 & 3.96150  \\
	$V_{s^*p}^{ac}$&2.52468& 2.96202 & 3.26633 & 2.11640 \\
	$V_{s^*p}^{ca}$&2.94540& 3.82679 & 1.86997 & 2.21680 \\
	$\lambda_a$&    0.32267& 0.36226 & 0.19373 & 0.14300  \\
	$\lambda_c$&    0.07567& 0.02717 & 0.01937 & 0.06700  \\
\end{tabular}
\end{table}
\noindent $^{\rm a}$ Reference [9]\\
$^{\rm b}$ Reference [31]\\
$^{\rm c}$ Reference [32]

\newpage


\begin{references}
\bibitem{bryant} G. W. Bryant, \prl\ {\bf 55}, 1786 (1985)
\bibitem{lowther} J. E. Lowther, J. Phys. C: Solid State Phys. {\bf 19},
1863 (1986)
\bibitem{kilday} D. G. Kilday, G. Maragaritondo, T. F. Ciszek, and 
S. K. Deb, J. Vac. Sci. Technol. B {\bf 6}, 1364 (1988)
{\bf 19}, 1863 (1986)
\bibitem{niles} D. Niles and H. H\"ochst, \prb\ {\bf 43}, 1492 (1991)
\bibitem{bala} W. Bala, M. Drozdowski, and M. Kozielski, Acta Physica 
Polonica A
{\bf 80}, 723 (1991)
\bibitem{gawlik} K. --U. Gawlik, J. Br\"ugmann, S. Harm, C. Janowitz,
R. Manzke, M. Skibowski, C. --H. Solterbeck, W. Schattke, and B. A.
Orlowski, Acta Physica Polonica A {\bf 82}, 355 (1992)
\bibitem{duc} T. M. Duc, C. Hsu, and J. P. Faurie, \prl\ {\bf 58}, 
1127 (1987)
\bibitem{arriaga} J. Arriaga and V. R. Velasco, Phys. Scripta {\bf 46},
83 (1992)
\bibitem{prb1} D. Olgu\'\i n and R. Baquero, \prb\ {\bf 50}, 1980 (1994)
\bibitem{rafa} R. Baquero, V. R. Velasco, and F. Garc\'\i a--Moliner, 
Physica Scripta {\bf 38}, 742 (1988)
\bibitem{noguera} R. Baquero and A. Noguera, Rev. Mex. F\'\i s. 
{\bf 35}, 638 (1989)
\bibitem{quintanar} C. Quintanar, R. Baquero, V. R. Velasco
and F. Garc\'\i a--Moliner, Rev. Mex. F\'\i s. {\bf 37}, 503 (1991).
\bibitem{rafa-prb} R. Baquero, A. Noguera, A. Camacho, and L. Quiroga,
\prb\ {\bf 42}, 7006 (1990)
\bibitem{angela} F. Rodr\'\i gez, A. Camacho, L. Quiroga, and R. Baquero, 
Phys. Status Solidi B {\bf 160}, 127 (1990)
\bibitem{daniel} D. Olgu\'\i n, M. Sc. Thesis, CINVESTAV-IPN, M\'exico 
1994.
\bibitem{prb2} D. Olgu\'\i n and R. Baquero, Phys. Rev. B {\bf 51}, 
16981 (1995).
\bibitem{g-m} F. Garc\'\i a--Moliner and V. R. Velasco, Prog. Surf. 
Scie. {\bf 21}, 93 (1986)
\bibitem{martin} R. Baquero, M. Ya\~nez, and M. Salmer\'on, J. Phys.
(Condens. Matter) {\bf 5}, A161 (1993).
\bibitem{rafa-moliner} V. R. Velasco, R. Baquero, R. A. Brito-Orta, and 
F. Garc\'\i a--Moliner, Condens. Matter {\bf 1}, 6413 (1989)
\bibitem{brito} R. A. Brito--Orta, V. R. Velasco, and F. Garc\'\i a--Moliner,
Phys. Scripta {\bf 37}, 131 (1988); \prb\ {\bf 38}, 9631 (1988)
\bibitem{col} A. Camacho, D. Olgu\'{\i}n and R. Baquero, Rev. Col. F\'{\i}s.
{\bf 27}, 611 (1995).
\bibitem{keldysh} L. V. Keldysh, Zh. Eksp. Teor. Fiz. {\bf 47},
1515 (1965) [Sov. Phys. JETP. {\bf 20}, 1018 (1965)]
\bibitem{falco} C. M. Falco and I. K. Schuller, in {\it Synthetic Modulated
Structures}, edited by L. L. Chang and B. C. Giessen (Academic, New York
1985)
\bibitem{falicov} L. Falicov and F. Yndurain, J. Phys. C: Solid St. 
Phys. {\bf 8}, 147 (1975)
\bibitem{sancho} M. P. L\'opez--Sancho, J. M. L\'opez--Sancho, and
J. Rubio, J. Phys. F: Metal Phys. {\bf 14}, 1205 (1984); {\bf 15}, 
855 (185)
\bibitem{trieste} R. Baquero, unpublished
\bibitem{slater} J. C. Slater and G. F. Koster, Phys. Rev. {\bf 94}, 
1498 (1954)
\bibitem{austria} P. Volg, H. P. Hjalmarson, and J. D. Dow, J. Phys. 
Chem. Solids {\bf 14}, 365 (1983)
\bibitem{harrison} W. A. Harrison, \prb\ {\bf 24}, 5835 (1981)
\bibitem{chadi} D. J. Chadi, \prb {\bf 16}, 790 (1977)
\bibitem{bertho} D. Bertho, D. Boiron, A. Simon, C. Jouanin, and 
C. Priester, \prb\ {\bf 44}, 6118 (1991)
\bibitem{ang-cdse} A. Camacho, private communication.
\bibitem{pelhos} K. Pelhos, S. A. Lee, and Y. Rajakarunanayake, \prb
{\bf 51}, 13 256 (1995); Y. Rajakarunanayake, R. H. Miles, G. Y. Wu, 
and T. C. McGill, \prb\ {\bf 37}, 10 212 (1988)

\end{references}
\end{document}